\documentclass[conference]{IEEEtran}

\makeatletter
\def\ps@headings{%
\def\@oddhead{\mbox{}\scriptsize\rightmark \hfil \thepage}%
\def\@evenhead{\scriptsize\thepage \hfil \leftmark\mbox{}}%
\def\@oddfoot{}%
\def\@evenfoot{}}
\makeatother
\usepackage{cite}
\usepackage[T1]{fontenc}
\usepackage{bm}
\usepackage{amssymb}
\usepackage{dsfont}
\usepackage{mdwlist}
\usepackage{graphics}
\usepackage{graphicx}
\usepackage{longtable}
\usepackage{amsmath}
\usepackage{mathtools}
\usepackage{multirow}
\usepackage{amsmath}

\usepackage{dsfont}
\usepackage{times}
\usepackage{subfigure}
\usepackage{epsfig}
\usepackage{latexsym}
\usepackage{amsfonts}
\usepackage{amssymb}
\usepackage{paralist}
\usepackage{comment}
\usepackage{xspace}
\usepackage{mathrsfs}
\usepackage{amssymb}
\usepackage{color}
\usepackage[ruled]{algorithm}
\usepackage[noend]{algorithmic}

\usepackage{amssymb}
\usepackage{url,epsfig,array}
\usepackage{leftidx}
\usepackage{amsmath}

\newcommand{\eqqref}[1]{Eq.~(\ref{#1})}

\def\ie{\textit{•}{et al.}\xspace}

\def\eg{\textit{e.g.}\xspace}

\begin{document}

\title{Rejecting the Attack: Source Authentication for Wi-Fi Management Frames using CSI Information}

\author{\IEEEauthorblockN{Zhiping Jiang\IEEEauthorrefmark{1}, Jizhong Zhao\IEEEauthorrefmark{1}, Xiang-Yang Li\IEEEauthorrefmark{2}, Jinsong Han\IEEEauthorrefmark{1}, Wei Xi\IEEEauthorrefmark{1}}
\IEEEauthorblockA{\IEEEauthorrefmark{1}School of Electronic and
  Information Engineering, Xi'an Jiaotong University, Xi'An, China}
\IEEEauthorblockA{\IEEEauthorrefmark{2}Department of Computer Science,
  Illinois Institute of Technology, Chicago, IL}
Email: \{\emph{jiangzp.cs, weixi.cs}\}@gmail.com, \emph{xli}@cs.iit.edu, \{\emph{zjz,hanjinsong}\}@mail.xjtu.edu.cn}

\maketitle

\begin{abstract}

Comparing to well protected data frames, Wi-Fi management frames (MFs)
 are extremely vulnerable to various attacks.
Since MFs are transmitted without encryption or authentication,
attackers can easily launch various attacks by forging the MFs.
In a collaborative environment with many Wi-Fi sniffers, such attacks can be easily detected by sensing the anomaly RSS changes.
However, it is quite difficult to identify these spoofing attacks without assistance from other nodes.

By exploiting some unique characteristics (\eg,
 rapid spatial decorrelation, independence  of Txpower, and much
 richer dimensions)  of 802.11n Channel State Information
 (CSI), we design and implement \emph{CSITE}, a prototype system to authenticate the Wi-Fi management frames on PHY layer merely by one station.
Our system CSITE, built upon \textit{off-the-shelf} hardware,
 achieves precise spoofing detection without collaboration and
 in-advance fingerprint.
Several novel techniques are designed to address the challenges
 caused by user mobility and channel dynamics.
To verify the performances of our solution, we conduct extensive evaluations in various
 scenarios.
Our test results show that our design significantly outperforms the
 RSS-based method. We observe about $8$ times improvement by CSITE over RSS-based method
 on the falsely accepted attacking frames.

\end{abstract}

\section{Introduction}
\label{sec:introduction}

Wi-Fi technology is on its rapid evolution. IEEE 802.11n and its successor 802.11ac supports more than
600Mbps throughput. 802.11i amendment, or WPA2 encryption, provides safe data exchange. However, an attacker  can still launch Denial of Service (DoS) attacks\cite{bellardo2003802,yang2006scan,pelechrinis2010denial} easily, breaking the connection between AP
and client, establishing rogue AP, and even performing
Man-In-The-Middle (MITM) attack.
To varying degrees, all these attacks exploit a main vulnerability of the 802.11 system, that the
 Management Frames (MFs), which are indispensable to the normal operation of Wi-Fi, has not been protected by any security measures ~\cite{arbaugh2002your}. Hence, attackers can forge the MFs simply using a laptop with an injection-enable wireless NIC.

Sequence Number (SN) based spoofing detection can be bypassed if the SN of attacking frames follows the original pattern. IEEE 802.11w amendment seeks to protect several key MFs by encryption-based authentication, it still has some vulnerabilities identified in recent researches~\cite{ahmad2011short,80211winfocom}.
The Mac address spoofing attacks can be easily detected by Wireless Intrusion
 Detection System (WIDS) or similar
 systems~\cite{sheng2008detecting,chumchu2011new,yang2012detection}.
 The power of WIDS roots on the collaboration of many Wi-Fi sensors dispersed in the environment.
These sensors overhear the Wi-Fi traffic and cooperate in
 detecting the anomaly Received Signal Strength (RSS) variation for the same MAC addresses. However, due to its high deployment cost, WIDS is not common for public environment.
Because of its high correlation with transmit power (Txpower) and distance, RSS is naturally more suitable for localization ~\cite{bo2012locating,xi2010locating,liu2010location,wu2012will} rather than spoofing detection. Hence, without collaboration from other nodes, RSS-based spoofing detection can be bypassed by Txpower scanning.

In search of a MFs authentication mechanism which supports operating independently on a single station,
we focus on the 802.11n PHY-layer information, Channel State Information (CSI), which is a large complex-number matrix that reveals the Channel Frequency Response (CFR) for each subcarrier of the underlying 802.11a/g/n OFDM system. CSI has some unique advantages, \eg, rapid spatial decorrelation, independence of Txpower, and rich data dimensions. After some proof-of-concept experiments, we believe CSI is an ideal alternative to RSS-based spoofing detection.

Based on these advantages, we design \textbf{CSITE}, a CSI-based management frame
authentication system using \textit{off-the-shelf} NICs.
The idea is simple yet effective: regardless of the frame type, data or management frames
 the transmission between AP and legitimate stations undergoes
 the same channel fading. Consequently, their CSI are exactly the same.
If an attacker injects a forged MF, the CSI of this frame
 must be quite different to the CSI trend learnt from previously
 accepted frames, thus we consider that this frame is suspicious.

Our CSITE system is based on a reasonable security assumption that the
 data frames with strong encryption, \eg, WPA2 under AES encryption
 with a strong password, is very hard to be cracked in a relatively short time~\cite{mitchell2005security}.
As a result, an attacker cannot forge a data
 frame that can be correctly decrypted by legitimate
 stations, hence the encrypted data frames which are correctly received and decrypted are considered to be
 sent from genuine stations, and the CSI of these frames are deemed to be the fingerprint of the wireless channel between genuine stations.

Despite an elegant solution, there are three main challenges that should be
 carefully addressed:

First, compared to one-dimensional temporal RSS data, the CSI information for each frame is a
 large complex number matrix of the size $N_{tx}\times
 N_{rx} \times 30$, where $N_{tx}$ and $N_{rx}$ denote the number of
 transmitting and receiving antennas respectively. Learning the CSI pattern and
Identifying anomaly data points in such a high-dimensional data stream
are big challenges when the frame receiving rate $f_r$ is high.

Second, it is the spatial decorrelation that makes CSI
 unforgeable, but this also makes the authentication intolerable to frequently-happened channel dynamics, \eg, those caused by crowd flow or user mobility.
In such conditions, there are inevitably some genuine MFs that are rejected.
 A mechanism should be carefully devised to
 guarantee the delivery of every genuine MF.

Third, transmitting all frames in "HT" rate is allowed in 802.11n Sepc.,
and it is indispensable for measuring CSI.
However, most MAC layer implementation still use the 802.11a/g compatible code which transmits the MFs in \textit{legacy} rate.

In our work, an accurate and efficient CSI-based attacking detector is first designed. Since MF only occupies a small portion of normal traffic, high accuracy checking can be achieved without affecting network throughput.
To cope with the channel dynamics, we devise a
 method called "CSI Resolution Enhancement" (CRE) to ensure the
 transmission of legitimate MF even under highly intensive channel
 dynamics.
Since it is standard-permitted and technically possible for sending
 management frames in "HT" rate,
 we modify the MAC layer implementation
 and  NIC drivers to enable the transmission of management frames in
 "HT" rate.

In summary, the main contributions of this paper are as follows.
We design CSITE, a cross-layer system based on CSI to perform
  PHY-layer source authentication for Wi-Fi management frames.
In addition to  the natural advantages of single-station accurate
 authentication,   CSITE can also cope with user mobility, and no
 cooperation or in-advance fingerprint is required.
We implement a prototype of CSITE using the off-the-shelf hardware and
 conduct extensive studies on the performance of our method in various
 scenarios.
Our evaluations show  that CSITE has excellent performance on
 accuracy, robustness, and efficiency.
It significantly outperforms the RSS-based method in the same
scenarios.
For example, when the client  and the attacker are
 walking with regular speed, CSITE
 accepts  some attacking frames with probability about $2\%$, while
 RSS-based method accepts attacking frames with probability about
 $18\%$ for the same scenario.
When only the client is moving, we observe similar improvement (about
 $8$ times) by CSITE over RSS-based method on the falsely accepted
 attacking frames.
A more significant improvement is obtained in stationary scenarios, see Section~\ref{sec:evaluation} for details.
To the best of our knowledge,
 we are the first to exploit the unique characteristics (\eg, rapid spatial
  decorrelation, independence to Txpower, and rich dimensions)
  of \emph{off-the-shelf} platform's Channel State Information (CSI)
  for authenticating management frames in Wi-Fi networks.

The rest of the paper is organized as follows.
Section~\ref{sec:related} presents some preliminaries and reviews
related works.
Section~\ref{sec:design} describes the CSITE system design.
A series of experimental results
 and analysis are shown in Section~\ref{sec:evaluation}.
We discuss the compatibility and other security issues in
Section~\ref{sec:discussion} and conclude the paper in
Section~\ref{sec:conclusion}.

\begin{figure}[t]
\begin{center}
\includegraphics[scale=0.75]{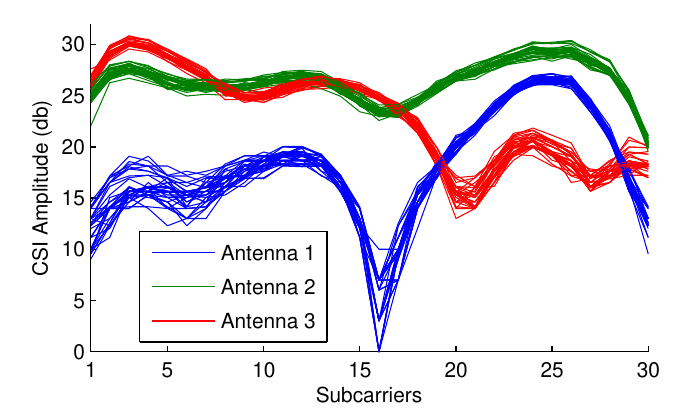}

\caption{An example of CSI data. Since Intel 5300agn NIC has 3
  antennas, there will be 3*30 subcarriers information for each MAC
  layer frames.  For visual clarity, here we show only 40
  frames. }
\label{fig:csisample}
\end{center}
\vspace{-0.1in}
\end{figure}

\section{Background and Related Work}
\label{sec:related}

In this section we first give a brief review of OFDM, CSI, and 802.11n, which are the foundations of
CSITE design. Then a review of related works are presented.

\subsection{OFDM, 802.11n, and CSI}

802.11a/g/n adopt Orthogonal Frequency Division Multiplexing (OFDM)
technology. In OFDM, the overall wide bandwidth channel is divided
into many small but orthogonal sub-carriers. Thus for OFDM system,
channel estimation is equivalent to measuring the parameters
of all the subcarriers. In 802.11n and its successor 802.11ac, Channel
State Information (CSI) is a large complex-number matrix which describes
the channel frequency response (CFR) for each subcarrier in every spatial stream.
Each complex value $h$ in CSI matrix
could be transformed to polar coordinates that

\begin{equation*}
   h=\left|h\right|{e}^{j\angle h}
\end{equation*}
where $\left|h\right|$ and $\angle h$ denote the amplitude and phrase of each subcarrier. Fig.~\ref{fig:csisample} presents the amplitude of CSI samples.



\subsection{Related Works}

Numerous researches claim to have the ability to detect MAC-layer spoofing attacks based on RSS or Sequence Number (SN).
 However, Txpower can be adjusted to forge the same RSS level,
 while SN could be forged by following the original pattern. Fingerprint
based on hardware transceiver profile is thought to be a perfect
solution~\cite{barbeau2006detecting}, but advanced attacker using
arbitrary waveform generator, can still compromise the
fingerprint~\cite{danev2010attacks}.

Wireless Intrusion Detection System (WIDS) or similar systems \cite{chen2007detecting,sheng2008detecting,
yang2009determining} can provide reliable attacking detection in secured environment, but these approaches are limited due to the deployment of monitor stations.
To the best of our knowledge, the most advanced RSS-based detection is the RCVI~\cite{zeng2011identity}. This work cleverly exploits the reciprocity of RSS variance in mobile wireless networks. By detecting the mis-matched RSS variation, an Identity-based Attack (IBA) is detected. However, RCVI require the sender to report the RSS records of the latest received ACK frames, which is a slightly high requirement.

There are growing interests in authentication, location distinction and
even localization based on physical layer information. Channel
Impulse Response (CIR) has been used to provide robust location
distinction in \cite{patwari2007robust,zhang2008advancing}. There are some works
\cite{sen2012spinloc,wu2012fila,sen2012spot} that went further trying to
provide precise indoor localization either by identifying the
Line-Of-Sight components or by identifying cluster information in
CSI.

A new attack against PHY-layer authentication called \textit{mimicry} was identified
in~\cite{liu12enhanced}. However, such attack is neither easy
to launch due to the existence of \textit{symbol sensor}, and it is not likely to
succeed due to the MIMO configuration which introduces richer channel
information.

\section{CSITE Design}
\label{sec:design}

In this section, we will first present some of our observations on
 which the design of CSITE are based. Then the design of CSITE is
 presented in details.

\subsection{CSI for Packet Authentication}
 CSI, in contrast to RSS, decorrelates with spatial position quite rapidly~\cite{WCbook}. The correlation efficient $\rho$ between two CSI samples may quickly drop to $0$ if the sampling locations are apart merely more than half a wavelength. It means once sufficiently distant, an attacker cannot estimate the victim's CSI based on the CSI measured locally.
Besides the decorrelation with position, CSI also has very low correlation with Txpower, therefore the traditional Txpower-scanning attacks cannot fool CSI-based detection.
 Third, CSI is a high-dimensional data~\cite{ieee2009ieee}. For a 3$\times$3 802.11n MIMO transferred frame, there are 270 values in the CSI matrix. Apparently it is of great difficulty to forge the CSI even
 under most sophisticated preparation.

\begin{figure*}[!t]
\begin{center}
\begin{tabular}{ccc}
\includegraphics[scale = 0.8]{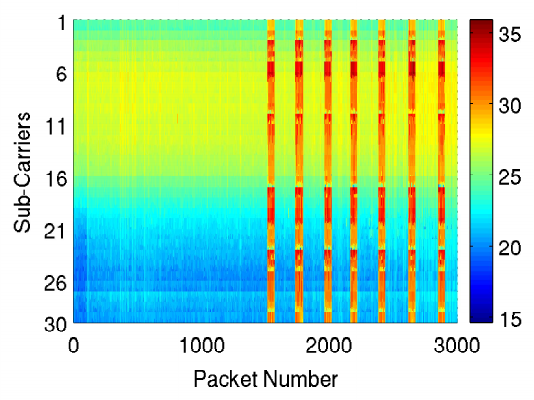} \hspace{-0.3in} &
\includegraphics[scale = 0.8]{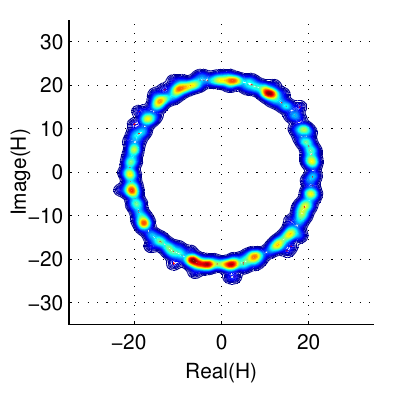} \hspace{-0.4in} &
\includegraphics[scale = 0.8]{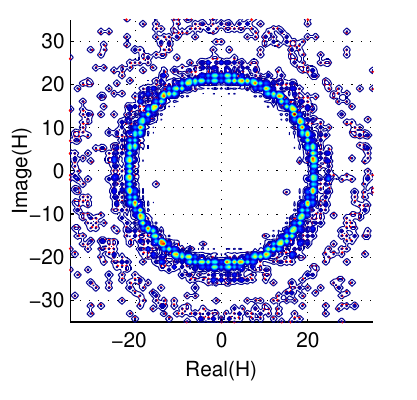} \\
(a) CSI sample with injection & (b) PDF for SC20 (first $1500$ frames) &
(c) PDF for SC20 (all data)
\end{tabular}
\end{center}
\caption{(a) Amplitude of CSI sample where warmer
  colors denote larger amplitudes. Attacking frames are injected
  starting from frame number $1500$.
(b)  PDF  for sub-carrier 20 of the first $1500$
   frames, where the value (denoted by colors) at a point is the
   number of frames with this CSI.
 (c)  PDF for  sub-carrier 20 of all samples, including attacking
   frames.}
\label{fig:injectionSample}
\vspace{-0.15in}
\end{figure*}

A simple attack experiment is conducted to observe the characteristics of CSI.
We collected 3000 frames, among which the first 1500 frames are from legitimate station, while the following 1500 frames include frames
 from both attackers and legitimate station. Fig.\ref{fig:injectionSample} (a) shows the
 CSI amplitude of the sample packets, where different colors denote
different amplitudes. We can see the attacker starts
injecting a group of attacking frames periodically after the 1500th frame, and the visual
difference between the legitimate and injected frames is very clear.
Fig.\ref{fig:injectionSample} (b) shows the empirical probability density function (PDF) of subcarrier 20 collected from the first $1500$ frames by directly plotting the $Re(h)$ and $Im(h)$ dots, where
 $h$ is the CSI for this subcarrier. Since the amplitude is stable during the test
 and the phase is distributed between $0$ to $2\pi$, we observed a ring-shaped structure with narrow width. Fig.\ref{fig:injectionSample} (c) shows the PDF of all the CSI for subcarrier 20. Due to the amplitude difference between the legitimate and injected frames, a double-ring shaped structure is presented. Such amplitude difference can be used to detect the attacking frames.

\subsection{CSITE Architecture}
\label{subsec:working-model}

We are now ready to discuss the architecture of our
  \textit{CSITE} prototype for authenticating MF frames in Wi-Fi
  environment.

The CSITE system consists of two parts: \emph{CSITE filter}
 and \emph{MF transmission assurance system}.
The CSITE filter implements our CSI-based
 spoofing detection algorithm,
 and its goal is to detect and reject
 any suspicious MFs.
 However, the safety and efficiency are always contradictory.
In dynamic environment, CSITE
 filter may also reject some legitimate MFs.
In such case, the sender
 should take measures to ensure the successful delivery of legitimate
 MFs without compromising the security standard of receiver's CSITE filter, and
 this is achieved by the MF transmission assurance system.

Since routine data frames are naturally used to update CSI pattern, we
don't exert extra burden to network traffic. However to cope with the
burst of transmission and asymmetry between uplink and downlink, we
set a maximum interval $T_{im}$ between two CSI updates.
Once a station has not been updated for a time duration exceeding
 $T_{im}$, it will send a ICMP "Probe Request/Reply" probe
 to force a CSI probe.
Then the update frequency, denoted as
 $f_s$, of a station would be $max(1/T_{im},f_{dl})$, where $f_{dl}$
 denotes the downlink data frames frequency.

\subsection{CSITE filter}
\label{subsec:filter}

The mission of CSITE filter is quite clear.
Let $S_{Y}$ denote the frame stream received by a station, and
$S_{Y}$ is composed of three parts: $ S_Y=\{S_{d}, S_{m}, S_{in}\}
$. Here $\{S_{d}\}$ and $\{S_{m}\}$ are the encrypted data frames
stream and management frames stream sent from genuine station,
respectively. $\{S_{in}\}$ is the forged frames stream sent by
attackers using injection tools. Our mission is to determine, for each
newly arrived management frame $M$,  whether it's sent from genuine
station or attacker based on the CSI pattern learnt from $S_{d}$.
On designing such an filter, there are two technical requirements:

$\bullet$ \textbf{Low False Positive (FP) error:}
Classifying frames into ``legitimate'' or ``suspicious''
 frames may introduce two errors: wrongfully
 accepting an attacking frame (called false negative (FP) hereafter), and
 wrongfully rejecting a legitimate
 frame (called false Negative (FN) hereafter).
Since re-transmission can be launched once a
   delivery fails, the FN error is tolerable to some extent.
However, due to the high risk of successive attacks (\eg,
 man-in-the-middle attacks) triggered by some
 spoofing attacks, such as de-authentication attack, the FP error
  is absolutely not acceptable.

$\bullet$ \textbf{Low overhead:}
 In a real world environment, network throughput could be very high,
 large computation and communication overhead for attacking detection
 will significantly degrade the network performance.

 Due to the rapid spatial decorrelation, the CSI of spoofing frames are
  highly probable to be "distant" from the CSI of legitimate frames.
 Thus, detecting spoofing frames can be viewed as an online
  anomaly detection problem and the goal is to identify such "distant
  points".

\textit{K-Nearest Neighbor} (KNN) \cite{hautamaki2004outlier} is a common solution for
high-dimensional anomaly detection \cite{aggarwal2001outlier}.
Notice that because a MF frame is said to be suspicious if it
 significantly deviates from the trend of \emph{most recently} accepted
 frames, the ``anomaly'' detection for the problem studied in this paper
 also needs to consider the temporal distance of the frames.
Thus, traditional KNN algorithms cannot be directly applied here.
To reflect the impact of the timing characteristics of all frames, our distance metric takes both spatial and temporal distance into account. An self-adaptive threshold is determined to classify a CSI point into two categories "trusted" or "suspicious".
However, before introducing the algorithm, we should first reduce the data point dimension.

\medskip

\textit{Dimensionality Reduction}:
Due to the high dimensions of CSI data point, it will consume large computational resource to perform anomaly detection if all dimensions are taken into consideration.
Even we set the MIMO Tx-antenna $N_{tx}=1$ and Rx-antenna $N_{rx}=3$ for
 our prototype, the dimension of the CSI data point for each frame is  $Dim(S_H)=N_{tx}\times N_{rx}\times 30=90$,
 which is still too large, especially for AP, which is going to handle multiple connections.

As \textit{phase} is distributed between $0$ to $2\pi$ which provides no discriminative information, the complex number data
point $H$ is first reduced to a real number data point containing only
the amplitude $ A=\vert H \vert$.
Since amplitudes of subcarriers exhibit certain continuous structure as shown in Fig.\ref{fig:csisample},
 we can further merge the adjacent amplitudes.
In our system, every 2 adjacent amplitudes are merged to their mean as $(A_i
 +A_{i+1})/2$.

\medskip

\textit{Frame Authenticity Verification:}
To verify the claimed authenticity,
 each receiver holds a sliding window $W_r$
 to store the latest verified CSI points with a length $L_{W}$.
Determining whether a MF is from genuine station is equivalent to
 determining how distant a MF is from to the CSI trend in $W_r$.
If the incoming MF frame perfectly follows the trend, it is highly
 likely to be a true MF;  otherwise it is \emph{suspicious}.
We will use the "\emph{degree of following}" (DoF) (exact definition
will be given later) to characterize how closely   a newly received MF
 $M$ follows the trend defined by frames in the sliding window $W_r$.
This DoF is determined by two factors: the distance to its $k$ nearest
 points in the window $W_r$  and the time difference between $M$'s arrival
 time $t_M$ and the  arrival time of its $k$ nearest points.

Suppose there are $n$ dimensions in each data point after
 \textit{dimensionality reduction}.
We first define the Euclidean distance between CSI point $A$ and $B$ as
 $dist(A,B)=(\sum_{i=1}^{n}(A_i-B_i)^2)^{\frac{1}{2}}$.
 The \textit{following coefficient}
 between these two points is defined as follow:

\begin{equation*}
fc(A,B)=e^{\lambda(|t_A-t_B|)}
\end{equation*}
where $\lambda$ is a constant called \textbf{time gain factor} and $t_A$
 denotes the arrival time of point $A$.
We then define the "time-gain distance" between point $A$ and $B$ as

\begin{equation*}
tgd(A,B) = dist(A,B) \mathbf{\cdot} fc(A,B)
\end{equation*}
Let $ N^{tgd}_k(M,W_r)=\{P_1, P_2, \cdots, P_k\}$ be the $k$-NN of $M$
 from the sliding window  $W_r$ under the TGD distance.
The "\emph{Degree of Following}" (\textit{DoF}) of a new arrival
 management frame $M$ is then defined as
\begin{equation}
DoF(M)=\dfrac{\sum_{i=1}^{k} tgd(M,P_i),}{k}|P_i \in N^{tgd}_k(M,W_r)
\label{eq:dof}
\end{equation}

\textit{Dynamic Threshold Scaling (DTS) :}
We use threshold $\tau$ to
 decide whether to accept a newly arrived frame $M$:
 the $M$ is considered to be legitimate \textbf{iff} $DoF(M)< \tau$.
Recall that the premier goal of CSITE is to
 prevent FP error, the $\tau$ should be adjusted adaptively  to defend attacks
 even under highly dynamic environment.
 Based on a reasonable assumption that the \textit{DoF} of a newly arrived legitimate MF $M$ is highly probable to be similar to the \textit{DoF}s of the recently accepted frames.
 Thus in our system $\tau$ is determined according to the latest DoFs.
 Let $Q_b(W_r)$ denote the most recently accepted $b$-th point in the
 window $W_r$.
Instead of using simple mean or median,
 the $\tau$ is set to $i$-th percentile of DoFs of recently accepted
  points, \ie
\begin{equation}
\tau = p_i(\{DoF(Q_b(W_r)) \mid 1 \le b \le k \})
\label{eq:thresh}
\end{equation}
where $p_i(S)$ denotes the $i$th percentile function. Although
\eqqref{eq:thresh} requires $k\times L_w$ calculation, it can be
optimized by pre-caching the distance matrix between points $P_i$ and $P_j$.

Here the right selection of percentile $i$ is vital for the system.
When there is little channel dynamics, average \textit{DoF}s of recently accepted frames could be very low. It may cause more FN error (reject the legitimate MF), thus slightly higher $i$ is preferred.
While there is intensive channel dynamics, average \textit{DoF}s of recently accepted frames could be very high. In such case the \textit{DoF} of a legitimate MF is not necessarily lower than the \textit{DoF} of an attacking frame, thus lower $i$ is preferred for security concerns. An negative correlation between $i$ and \emph{Channel Stability} is needed.

In CSITE, we define the channel stability $\sigma_W$ as the mean of the
standard variance of the differences between two adjacent CSI points that

\begin{equation*}
\sigma_W = \overline{(std_n(|P_j - P_{j+1}|))}, n\in[1,Dim(P_i)],j\in [1,L_w-1]
\end{equation*}
where $std_n$ stands for the standard variance for $n$th dimension of
the CSI in window.
We then define an effective negative correlation between $i$ and $\sigma_W$ as
$i_1 = \dfrac{i_0}{\sigma_W / \sigma^r_W}$,
where $i_0$ is set to 75 as default, and $\sigma^r_W$ denotes the
reference $\sigma_W$, which is measured during the CSITE
initialization.

Based on the definition of $DoF(M)$, $\tau$, and $i$, we design our source
authentication algorithm as shown in Algorithm.\ref{alg:detection}.

\begin{algorithm}[ht]
\caption{Spoofing Frame Detection Algorithm}
\label{alg:detection}
\begin{algorithmic}[1]
\REQUIRE \

The CSI amplitude of a newly received frame $H$;\

The encryption property of $H$, \\
 $attr_{en}(H)\in\{encrypted,unencrypted\}$
\ENSURE \
A security classification of $H$, \\
$attr_{sec}(H)\in\{trusted,suspicious\}$
\FOR{\textbf{each} new arrival frame $H$}
\IF{$attr_{en}(H) == encrypted$}
\STATE sliding window $W_r$ move forward to include $H$
\STATE $attr_{sec}(H)=trusted$
\ELSE
\STATE calculate the $DoF(H)$ according to eq.\ref{eq:dof}
\STATE calculate the $\tau$ according to eq.\ref{eq:thresh}

\IF{$DoF(H) \leq \tau$}
\STATE sliding window $W_r$ move forward to include $H$
\STATE $attr_{sec}(H)=trusted$
\ELSE
\STATE $attr_{sec}(H)=suspicious$
\ENDIF
\ENDIF
\ENDFOR
\end{algorithmic}
\end{algorithm}

\begin{figure*}[!t]
\begin{center}
\begin{tabular}{cccc}
\hspace{-0.1in}
\includegraphics[scale = 0.6]{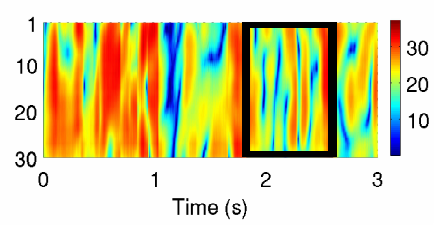} \hspace{-0.1in} &
\includegraphics[scale = 0.6]{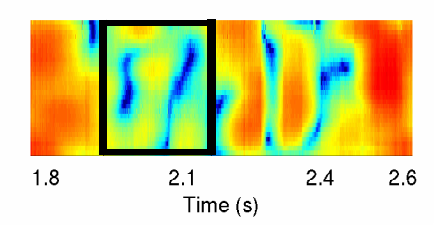} \hspace{-0.15in} &
\includegraphics[scale = 0.6]{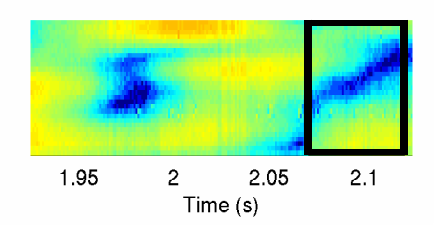} \hspace{-0.15in} &
\includegraphics[scale = 0.6]{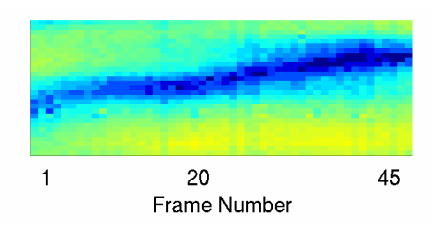} \\
(a) 100\% view & (b) 375\% zoomed view & (c) 1500\% zoomed view & (d) 6000\% zoomed view
\end{tabular}
\end{center}
\caption{(a). Original view of CSI under highly dynamic
  environment. (b), (c), and (d) provide gradually zoomed views for the details circled by black box. }

\label{fig:highresolution}
\vspace{-0.1in}
\end{figure*}

\subsection{MF transmission assurance system}
\label{subsec:assurance}

Due to the rapid spatial decorrelation and the negative correlation
between $i$ and $\sigma_W$,
the CSITE filter is more likely to reject than to accept any suspicious frames.
In dynamic environment, it is even harder to classify a MF into "trusted" due to the large noise.
How to guarantee the delivery of legitimate MFs in any case is a big problem.

Despite rapid spatial decorrelation, wireless signal propagation can be well modeled as
an analogue continuous system. In this system when sampling rate $f_s\rightarrow \infty$, the differences between each sampling $\Delta D\rightarrow0$.
It means when the frame rates is high enough, we can see very smoothed and slow-changing CSI amplitude surface under intensive channel dynamics.
Fig.~\ref{fig:highresolution} presents a proof-of-concept experiment.
During the experiment, large files are transmitted in HT rate between fast-moving stations, Fig.~\ref{fig:highresolution}(a) presents the temporal CSI data of a station. When
 we gradually zoom into the details specified by the black rectangle, we
 see very smooth surface just like in static environment.

Based on the observation, we design a method called "CSI Resolution Enhancement" (CRE)
to guarantee the delivery of MF. The core of CRE is that:
if we transmit an unprotected
MF $M$ immediately after a group of high frequency
"precursor" data frames,
there will be smoothed amplitude surface in $W_r$
and receiver's CSITE filter will think
it is in a static environment and accept the $M$ by setting higher $i$.
The sender repeats this procedure until the delivery succeed.

Formally speaking, for each MF $M$ to be transmitted,
we define a frame stream
 $S_j=\{D_0,D_1,...,D_{l_{j}},M\}$ with minimum
 transmission interval between frames, where $D_i$ are encrypted data frames.
 $S_j$ is the $j$-th transmission procedure.
 We repeat this procedure until the frame $M$ is successfully transmitted.

Since the proportion of MF in normal communication is small,
we adopt a simple yet robust power-based scheme to guarantee the delivery. Suppose
both sides keep the same $k$ and $L_W$, the $l_{j}$ is expected to be:
\begin{equation}
l_{j}=2^{j}\times (l_{1}+1), j \in(2,3...,N), l_{j} \le L_W
\end{equation}
This scheme simplifies the problem, and we only need to determine the
initial value $l_{1}$, which  is a fixed value in current prototype.


\textit{Negative ACK encapsulated in Echo Request:} There is a
firmware-level limit: we have no control on the transmission of
ACK frame. The firmware will emit the ACK frame even if the frame is rejected by the CSITE filter,
therefore transmitter cannot determine if the delivery is successful.

We adopt an \textit{ad hoc} solution to inform the transmitter. Every time
a frame does not pass the CSI filter, the receiver will immediately send
a \textit{Negative-ACK} to inform the sender. Such N-ACK is carried in a
ICMP "ECHO REQUEST" frame whose echo content indicates the
the failed frame type and
sequence number, like
"PROBE\_REQUEST@42316". Since the frame is encrypted, only genuine
transmitter can learn this N-ACK and start
re-transmission as described above. However if such N-ACK emits
for a spoofing attack, the genuine station will be aware of being
forged and may trigger alarm.

We mention again that this \textit{ad hoc} solution exists
only because we have no control on the ACK frames.

\section{Prototype Evaluation and Analysis}
\label{sec:evaluation}

\subsection{Prototype System Setup}
Our prototype system consists of 3 laptops equipped with Intel 5300 NICs. Two of them form an AP-Client network, and the rest one acts as an attacker. Their drivers are all modified to enable them to transmit (or inject) management frames in HT rate in compliance to IEEE 802.11n standard.

\subsection{Attacking Test Setting}

In order to fully evaluate the performance of CSITE filter and make comparison to RSS based detection, we designed test cases and applied them in 7 typical scenarios. The description of them is presented in Table \ref{tab:scenarios}. Scenarios A, B, and C test the performance when the client is stationary while channel dynamics are gradually increasing. D to F test the performance when client is moving with different speeds. G presents the last test that both client and attacker are moving.

\begin{table}[!h]
\caption{ Test Scenarios Description}
\label{tab:scenarios}
\begin{center}
\begin{tabular}{|c|p{6.5cm}|}
\hline
A & Both the client and attacker are stationary in a controlled environment.\\ \hline
B & Same as A, but there is some channel dynamics caused by crowd flow.  \\ \hline
C & The client is stationary, while the attacker is moving around. No crowd flow. \\ \hline
D & The client is moving, while the attacker is stationary. speed is normal. \\ \hline
E & The same as D, but moving speed is slow. \\ \hline
F & The same as D, but moving speed is fast.\\ \hline
G & Both the client and attacker are moving, speed normal. \\ \hline
\end{tabular}
\end{center}
\end{table}

We run a test for each scenario and each test lasts for 5 minutes. During the test, the AP and the client are continuously updating the CSI pattern using the \textit{ping} command. Besides the data stream generated by ping command, the client initiates 20 Probe Requests to the AP every 0.3s, and the AP replies 20 Probe Responses to the client immediately. Both Probe Request/Response are MFs and they form the un-encrypted stream $S_{m}$.

The attacker uses \textit{aireplay-ng} to inject 64 forged de-authentication frames to the client every 0.5s using the AP's MAC address. During the attack, the Txpower is scanning from 1dBm to 15dBm in a loop. For the sake of convenience, we set a switch in the client to prevent the connection being really de-authenticated once the client wrongfully accepts the forged de-authentication frames.

For each test case, we mainly focus on two error rates: FP error rate and FN error rate. Specifically, the FP error rate is the number of de-authentication frames which are considered to be sent from legitimate station over the totally received number of de-authentication frames. Similarly, the FN rate is the number of Probe Responses that are considered to be suspicious over the totally received number of Probe Responses.


\begin{figure}[hptb]
\begin{center}
\includegraphics[scale=0.85]{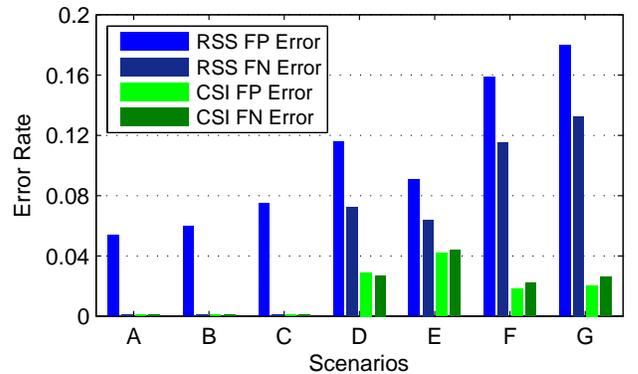}
\end{center}
\caption{Error rates comparison between CSITE and RSS-based detection }
\label{fig:RSSvscsihist}
\vspace{-0.1in}
\end{figure}

\subsection{Compared to RSS-based Authentication}


To make a fair comparison between CSITE and RSS-based detection, we turn off the dynamic threshold scaling (DTS) function and set the default value for $i=75$. Fig.\ref{fig:RSSvscsihist} presents the error comparison between CSITE and RSS-based detection in different scenarios. In stationary scenarios A, B, and C, due to the strong measurement noise, RSS-based detection yields an FP error rate about 6\%, while  CSITE achieves perfect 0 FP error rate. In motion scenarios, CSITE accepts only about 2\% attacking frames, while RSS-based detection accepts more than 17\% attacking frames. It is about 8 times improvement made by CSITE over the RSS-based detection.

\subsection{Impacts of various parameters}
To identify the impacts of various parameters, we still turn off the \textit{DTS} function and use default values $k=5$, $\lambda=1$, $L_w=40$, $i=75$ for the rest of evaluations if not specifically mentioned.

\begin{figure*}
\begin{center}
\begin{tabular}{cc}
  \hspace{-0.1in}\includegraphics[scale = 0.75]{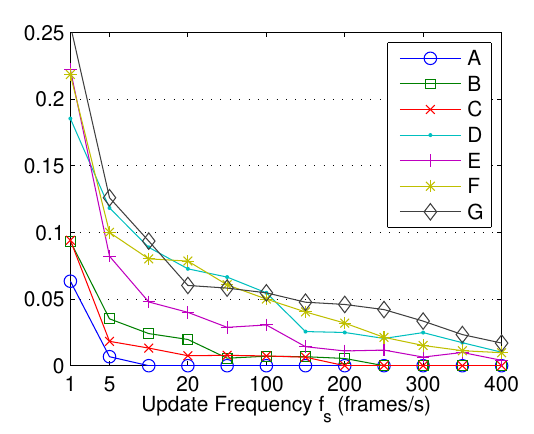}
   & \includegraphics[scale = 0.75]{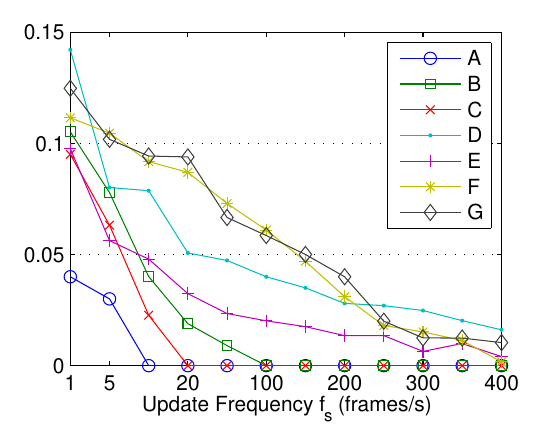} \\
  (a) FP error rate & (b) FN error rate
\end{tabular}
\end{center}
\caption{Impacts of update frequency on dete52ction error rate. Both the FP and FN errors decrease when update frequency $f_s$ is increasing. In Scenarios A, B, and C, both the error rate quickly converge to $0$, while for Scenarios D, E, F, G, higher frequency are needed to cut down the error rate. }
\label{fig:freqScanExp}
\end{figure*}

\begin{figure*}
\begin{center}
\begin{tabular}{cccc}
\hspace{-0.25in}\includegraphics[scale = 0.53]{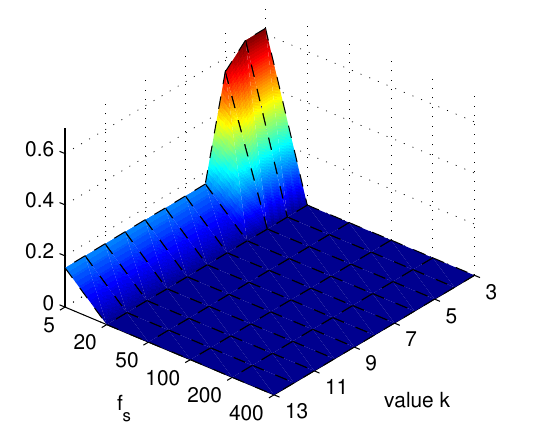} \hspace{-0.3in} &
\includegraphics[scale = 0.53]{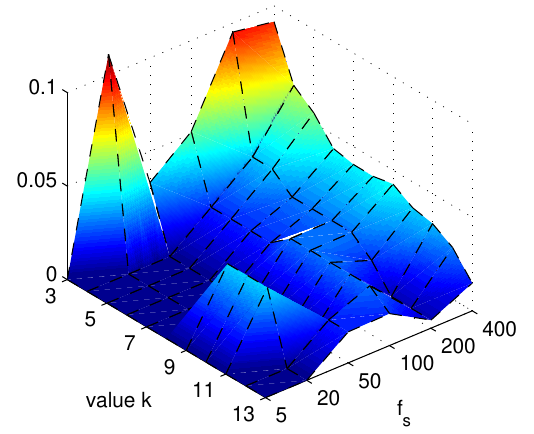} \hspace{-0.3in} &
\includegraphics[scale = 0.53]{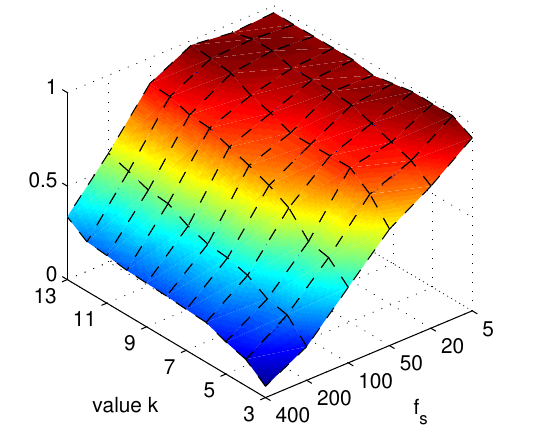} \hspace{-0.3in} &
\includegraphics[scale = 0.53]{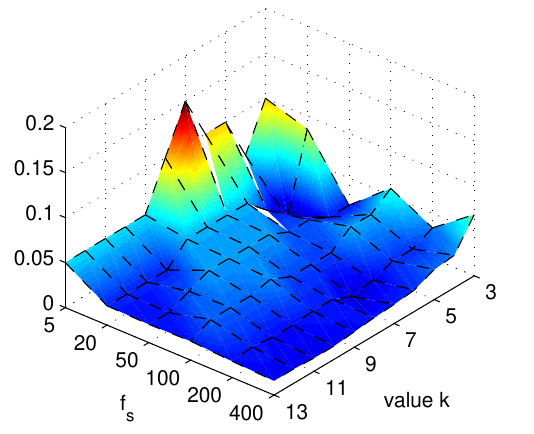}  \\
(a) FP error rate for A&(b) FN error rate for A&(c) FP error rate for G&(d) FN error rate for G
\end{tabular}
\end{center}
\caption{The combined impacts of value $k$ and update frequency $f_s$ in two scenarios A and G. (a) and (b) show the FP and FN error rates for Scenario A, and (c) (d) for G.
}
\vspace{-0.1in}
\label{fig:kscan}
\end{figure*}

\subsubsection{Impact of update frequency $f_s$}

To test the impact of $f_s$, we vary the sampling rate $f_s$ from $1$Hz to $400$Hz by uniformly dropping frames in the data stream. Fig.\ref{fig:freqScanExp} illustrates the FP and FN error rates in different scenarios when $f_s$ is increasing. In stationary scenarios, the FP and FN error rates drop to 0 rapidly when $f_s$ is increasing. For motion scenarios, the FP error rate drops to about 5\% when $f_s \geq 100$Hz. When $f_s \geq 400$Hz, the FP error rate is not higher than 3\% even under most intensive dynamics in scenario G.

 However, we should mention that in normal communication the \textit{DTS} function is turned on, the CSITE filter could reject almost all attacking frames even when $f_s$ is very low, as verified by our results in \textit{Impact of Dynamic Threshold Scaling}.

\subsubsection{Impact of the number of nearest neighbors $k$}

Fig.\ref{fig:kscan} shows the combined impact of $k$ and $f_s$ on the error rate in scenarios A and G. Since $\tau$ is partly determined by the \textit{time-gained distance} of the latest accepted points, $k$ and $f_s$ play an important role for deciding which points are taken into account.
According to Fig.\ref{fig:kscan}(a) and (b), higher $k$ could reduce the error rate when in stationary situation. In motion scenarios, however, lower $k$ is better. This is because higher $k$ in this case will introduce more non-related CSI points, which become noises when determining the $\tau$. Based on the test conducted in all scenarios, we believe $k=5$ is a suitable value for both the stationary and the motion scenarios.

\subsubsection{Impact of sliding window length $Lw$}

\begin{figure*}[!ht]
\begin{center}
\begin{tabular}{cccc}
\hspace{-0.25in}\includegraphics[scale = 0.54]{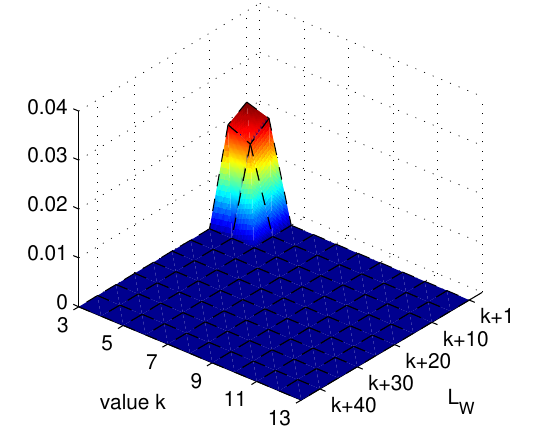} \hspace{-0.3in} &
\includegraphics[scale = 0.54]{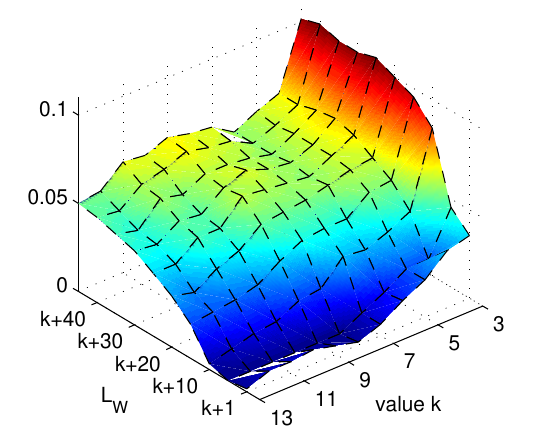} \hspace{-0.3in} &
\includegraphics[scale = 0.54]{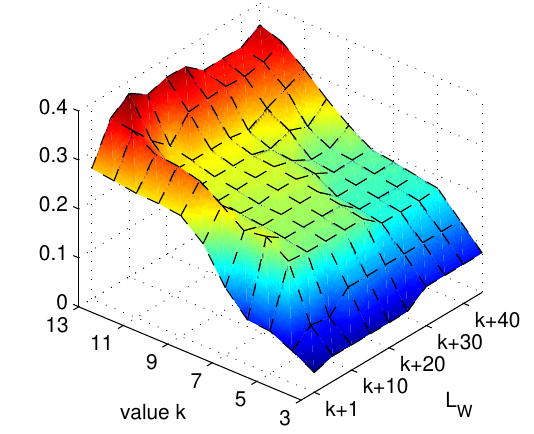} \hspace{-0.3in} &
\includegraphics[scale = 0.54]{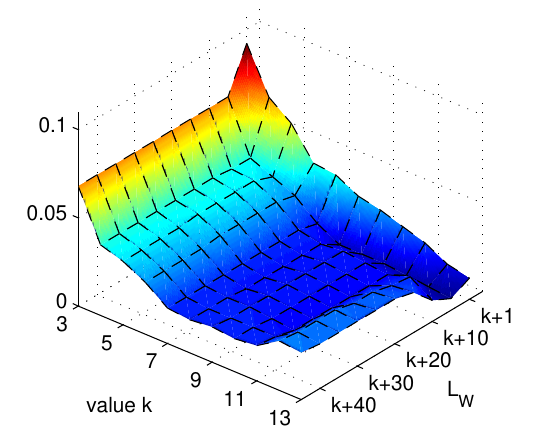}  \\
(a) FP error rate for A&(b) FN error rate for A &(c) FP error rate for G&(d) FN error rate for G
\end{tabular}
\end{center}
\caption{The combined effect of value $k$ and sliding window length $L_w$ on error rate under two scenarios A and G.}
\vspace{-0.1in}
\label{fig:WLScan}
\end{figure*}

Fig.\ref{fig:WLScan} presents the combined effect of sliding window length $L_W$ and $k$ on scenarios A and G. In both the stationary and the motion situations, increasing $L_W$ is generally good for reducing error rate, but the marginal effect is reduced since the CSITE filter is tuned to choose the most recently accepted points. When $ L_W > 40+k$, the benefit of increasing $L_W$ can be ignored in all scenarios, therefore we set $L_W = k+40$ for both accuracy and efficiency.

\subsubsection{Impacts of Dynamic Threshold Scaling}

\begin{figure*}[!ht]
\begin{center}
\begin{tabular}{cccc}
\hspace{-0.2in}
\includegraphics[scale = 0.65]{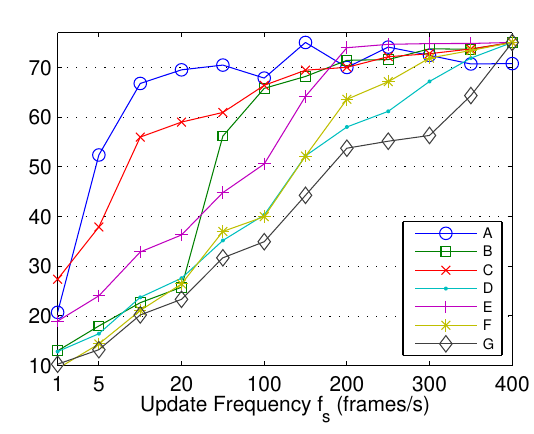}  & \hspace{-0.25in}
\includegraphics[scale = 0.65]{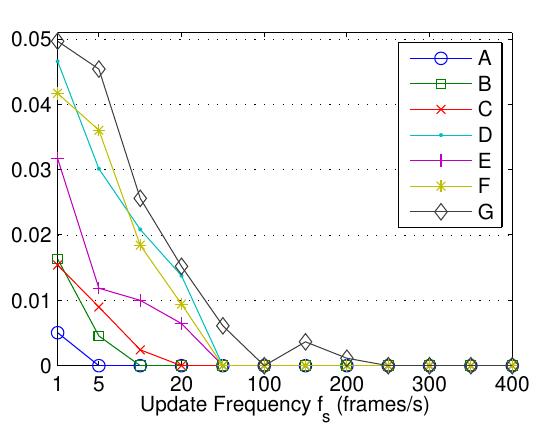}  & \hspace{-0.25in}
\includegraphics[scale = 0.65]{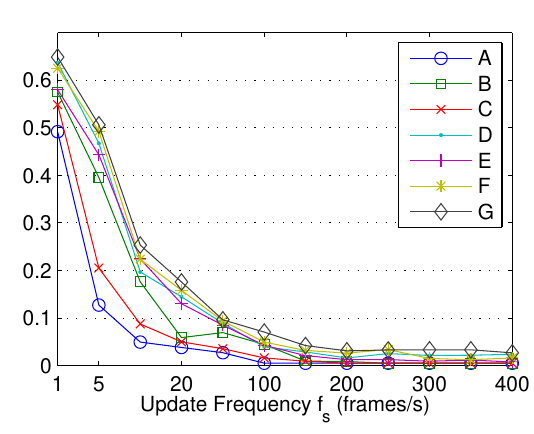}  \\
(a) Scaled Percentile $i$ &(b) FP error rate &(c) FN error rate
\end{tabular}
\end{center}
\caption{(a) shows the calculated percentile $i$ according to $\sigma_W$. (b) and (c) present the corresponding FP and FN error rates using the percentile $i$ shown in (a).}

\label{fig:optimization}
\end{figure*}

Apparently, lower threshold $\tau$ determined by $i$ rejects not only the attacking frames but also some legitimate frames which deviate from the trend center. However, a lower $i$ is preferred since the premier goal of CSITE is to reject the attacking frames. Recall the \textit{DTS} function introduced in Section III, Fig.\ref{fig:optimization}(a) shows the average window variance $\overline{\sigma_W}$ for different scenarios and frequencies. Fig.\ref{fig:optimization}(b) shows the dynamic percentile $i$ calculated according to $\sigma_W$. The impact to the error rate is shown in Fig.\ref{fig:optimization}(c) We see that the FP error rates quickly drops to $0$ in stationary scenarios. For the most dynamic scenario G, FP error rate drops to astonishingly 5\% when $f_s=5$Hz, and The FP error rates drop to 1.53\% when $f_s=20$Hz.

\subsection{Evaluation on MF transmission assurance system}
To fully evaluate the transmission of MF in different $f_s$ and scenarios, the data frames before the precursor frames are randomly dropped to simulate different $f_s$, and the length of precursor frames $L_{pre}$ varies from $0$ to $L_W$.

\begin{figure*}
\begin{center}
\begin{tabular}{ccc}
  \hspace{-0.1in}\includegraphics[scale = 0.79]{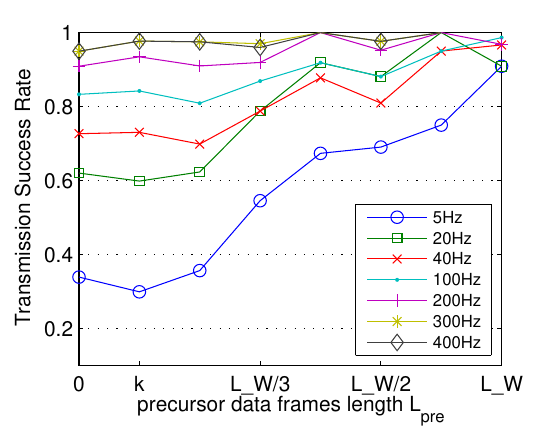}  &
  \includegraphics[scale = 0.79]{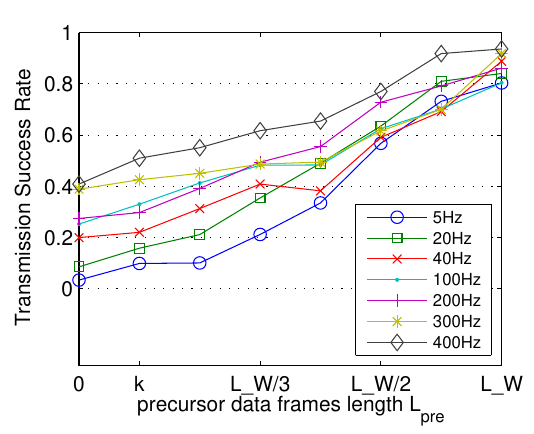} & \hspace{-0.3in}\\
  (a)  Successful Receiving Rate in A  & (b) Successful Receiving Rate in G 
\end{tabular}
\end{center}
\label{fig:transmit}
\vspace{-0.0in}
\end{figure*}

Fig.\ref{fig:transmit}(a) and (b) present the MF transmission success rate comparison with different amount of precursor frames in scenarios A and G. Since \textit{DTS} function is turned on, many FN errors are generated when $f_s$ is low. However, with the help of precursor frames, we can see that even when $f_s=5$Hz, with the length $L_{pre}=L_w$, 90\% and 78.5\% one-shot success (Transmission of $M$ succeeds with only one transmission) rate can be achieved in scenarios A and G. When $f_s=40$Hz, 97.3\% and 90.1\% one-shot success rate can be achieved. 

\section{Discussions}
\label{sec:discussion}

\textit{Driver Enhancement}:
In Intel IWL5300 NIC driver, there are some codes dealing with rate control for different situations.
In our prototype, the rate control is modified to transmit the MFs using the same MCS rate of latest successful HT transmission. If it fails (no ACK reply), both precursor data frames and MF will be transmitted in the lowest MCS value in the "BasicMCSSet" to ensure the success delivery. We mention again that these modifications are permitted according to the "\textit{Multirate Support}" in 802.11n Specification\cite{ieee2009ieee} Clause 9.6.

\textit{Source authentication for control frame}: Theoretically, if CSI value could be obtained for control frame, similar source authentication could be applied to control frames. However, it is currently impossible to achieve this goal, since the ACK feedback mechanism is hard-coded in firmware which is a binary file compiled from closed-source code.

\textit{Vulnerability of Man-In-The-Middle Attack}:
Since CSITE detects spoofing MFs based on the CSI of encrypted data frames, if the data frames are replayed in physical layer, Man-In-The-Middle (MITM) attack may succeed. To launch MITM, the attacker must be able to jam the paired stations and simultaneously tunnel their traffic through the attacker. To detect such attacks, users only need to open a virtual monitor interface. If these is the jamming, the overheard flows to different address will disappear simultaneously, which is impossible in normal situation. This mitigates the impact of MITM attack. However, we believe in most of attack scenarios, such kind of powerful attacker does not exist.

\section{Conclusion}
\label{sec:conclusion}

Management frame, the basis for operating 802.11 network normally, is extremely vulnerable to attacks.
Spoofing detection without cooperative information is unreliable using existing methods. Based on off-the-shelf hardware, we design CSITE, a Wi-Fi management frame source authentication system.
It leverages the unique characteristics of CSI to verify the authenticity of MFs, and the detection is tuned to be highly strict for False Positive (FP) errors.
To guarantee the successful delivery of MFs even under most intensive channel dynamics, we devise a method called CRE, which makes the MFs pass the detection with the help of precursor frames.
 Extensive evaluations are conducted to verify the performance of our system. These evaluations show excellent authentication ability and strong rejection against spoofing attacks.


\section*{Acknowledgement}
This work was supported in part by the NSFC Major Program under Grant 61190110, the NSFC under Grants 61228202, 61033015 and 61170216, China 973 Program under Grant No.2011CB302705, NSF CNS-0832120, NSF CNS-1035894, NSF ECCS-1247944, and the Fundamental Research Funds for the Central Universities of China.

\bibliographystyle{IEEEtran}
\bibliography{reference}

\end{document}